\documentclass[%
 reprint,
showpacs,preprintnumbers,
 amsmath,amssymb,
 aps,
 prd,
 lengthcheck,%
]{revtex4}
\usepackage{mathtext,amssymb,graphicx}
\usepackage{epsfig}
\usepackage{caption}
\usepackage[T2A]{fontenc}
\usepackage[utf8]{inputenc}
\newcommand{\be}{\begin{equation}}
\newcommand{\ee}{\end{equation}}
\newcommand{\nc}{\newcommand}
\newcommand{\tr}{\mathop{\mathrm{tr}}}
\nc{\bi}{\bibitem}
\newcommand{\ba}{\begin{array}}
\newcommand{\ea}{\end{array}}

\newcommand{\beq}{\begin{equation}}
\newcommand{\eeq}{\end{equation}}

\newcommand{\myref}[1]{~{(\ref{#1})}}
\newcommand{\mycite}[1]{~{\cite{#1}}}
\newcommand{\myfigref}[1]{~{Fig.~(\ref{#1})}}

\newcommand{\gev}{\mbox{GeV}}
\newcommand{\dirac}[1]{\lefteqn{#1}/\,}

\makeatletter
\newcommand\figcaption{\def\@captype{figure}\caption}
\makeatother

\begin{document}
\topmargin -0.8cm
\preprint{ITEP-TH-57/09}
\title{Quark Condensate and Effective Action from Dyson--Schwinger Equations}

\date{December 9, 2009}

\author{A.~V.~Zayakin$^{1,2}$}
 \author{V.~Khandramai$^{3}$}
\author{J.~Rafelski$^{4,1}$}

\affiliation{$^1$
Department f\"ur Physik der Ludwig-Maximillians-Universit\"at M\"unchen und\\
 Maier-Leibniz-Laboratory, Am Coulombwall 1, 85748 Garching, Germany
}
\vspace*{2mm}
\affiliation{$^2$
ITEP, B.Cheremushkinskaya, 25, 117218, Moscow, Russia}
 \vspace*{2mm}
\affiliation{$^3$
ICAS, Gomel State Technical University, Gomel 246746, Belarus}
\vspace*{2mm}
\affiliation{$^4$  Department of Physics, University of Arizona, Tucson, Arizona, 85721 USA}

\vspace{1cm}

\begin{abstract}
We obtain the  QCD quark condensate  from consideration of unquenched
quark dynamics in Dyson-Schwinger gluon vacuum. We consider the non-local
extension of the condensate and determine the quark virtuality. We also obtain
the condensate-driven contribution of the non-perturbative QCD 
to Euler--Heisenberg Lagrangian of QED in external electromagnetic fields.
\end{abstract}
\pacs{12.38.Aw,12.38.Lg,11.15.Tk}

\maketitle

\section{Overview}
We describe a  method relying on   Dyson--Schwinger equations (DSE)
for  obtaining (non-local) quark condensate and Euler-Heisenberg type effective action
with quarks in loops.  We develop a self-consistent scheme based on a full set of
DSE with dynamical quarks, ghosts and gluons. Our approach is built on
methods and prescriptions we adapt from  Fischer\mycite{Fischer:2003zc}.

Non-local quark condensate was first introduced by Mikhailov and Radyushkin~\mycite{Mikhailov:1986be},
and further developments followed soon after~\mycite{Mikhailov:1988nz,Mikhailov:1991pt}. 
The gauge invariant  NLC is defined by
\beq
C(x^2)\equiv\langle \bar{q}(x)E(x;0)q(0)\rangle,
\eeq
where Wilson phase factor is defined as
\beq
E(x;0)=\mathop{\mathrm{Pexp}}\left(ie\int_\mathcal{C} A_\mu(x)dx^\mu\right),
\eeq
and the contour $\mathcal{C}$ connects points $x$ and $0$.
In this paper we focus our attention on the first terms in powers of $x$ which are
independent of the Wilson line contributions. Wilson line terms are in general very
important, and  such contributions should be evaluated
self-consistently~\mycite{Semenoff:2002kk}, which may be done
by the Ericson-Semenoff-Szabo-Zarembo (ESSZ) technique
used by us in the previous work~\mycite{Zayakin:2009jz}. It has been shown
within the instanton vacuum model~\mycite{Trevisan:2004pj}
that the form of the NLC is nearly  independent
on possible irregularities of the path,  such as a cusp and thus in general the
path can be represented by a straight line.

The initial motivation for introducing a NLC came from
its influence on hadron phenomenology. For this reason NLC
has been  decomposed into the local condensates (LC) and the measure of the quark fluctuations
in vacuum, known as the quark virtuality (QV). This quantity related to NLC,   is defined as
\beq
\lambda_q^2=\frac{\langle \bar{q}D_\mu^2 q\rangle}
{\langle\bar{q}q\rangle},
\eeq
(here $D_\mu$ is the covariant derivative), arising in the standard operator product expansion
(OPE) of the NLC as the coefficient in front of the quadratic term:
\beq
C(x^2)
=\langle\bar{q}(0)q(0)\rangle
\left[1+\frac{x^2}{4}
\frac{\langle \bar{q}D^2 q\rangle}{\langle \bar{q}q\rangle}\right].
\eeq
Quark virtuality is related to the gluon-quark trilinear (local) condensate
\beq
\frac{\langle \bar{q}D^2 q\rangle}{\langle \bar{q}q\rangle}\sim
         \langle\bar{q}g\sigma_{\mu\nu}G^{\mu\nu} q\rangle.
\eeq
and thus can be counted as an independent vacuum structure parameter, characterizing
the non-perturbative QCD vacuum.
The standard estimate for $\lambda_q^2$ by Chernyak and Zhitnitsky~\mycite{Chernyak:1983ej} is $\lambda_q^2\approx 0.4\pm 0.1 \mathrm{GeV}^2$. There are larger estimates however, e.g. Shuryak suggests~\mycite{Shuryak:1988kd} $\lambda_q^2\sim 1.2\mathrm{GeV}^2$. We note that these numerical values for the  correlation length
are comparable with the typical hadronic scale.

Our effort to relate DSE and NLC
is not the first. An attempt to derive self-consistent equations
upon condensates was made by Pauchy Hwang~\mycite{Hwang:1997iq} in the large-$1/N_c$ limit;
unfortunately, this was not developed further. Non-local quark condensate
has been obtained within the flat-bottom potential approach to Dyson-Schwinger
equations, where typical correlation length of $3 \mathrm{GeV}^{-1}$
has been obtained~\mycite{Wang:2001gn}. Dyson--Schwinger equations
are solved in~\mycite{Zhou:2003bv} for quark dynamical mass and wave-function
(no gluons or ghosts solved dynamically; gluon propagator mimicked by an Ansatz,
rainbow approximation applied to quark equations); using propagators, the quark-quark
non-local condensate and the quark-quark-gluon local condensates are calculated,
typical correlation length obtained is $0.5 \mathrm{GeV}^{-1}$. Same methods were
used in~\mycite{Zhou:2004hx}, where virtualities 
$\lambda^2_{u,d}=0.7\mathrm{GeV}^2,
\lambda^2_{s}=1.6\mathrm{GeV}^2$ are reported. These results are confirmed
in\mycite{Zhou:2008zzl} and completed with gluon virtuality as well
$\lambda^2_{g}=0.2\alpha_s^{-\frac{1}{2}}-1.0\alpha_s^{3} \mathrm{GeV}^2$,
the latter exhibiting strong scale dependence via coupling constant.
Dyson--Schwinger equations were solved in a similar approximation
 (no dynamical gluons and ghosts) in\mycite{Zhou:2009ms}; however,
surprisingly large values of virtualities have been reported:
 $\lambda^2_{u,d}=12\dots 16\mathrm{GeV}^2, \lambda^2_{s}=14\dots 18\mathrm{GeV}^2$.
Till now, there has been no self-consistent treatment of the non-local condensates
based on DSE with gluons. We consider this to be a disadvantage of the scheme,
since quark fluctuations in vacuum are driven by gluons. We will present our result for QV as
function of quark mass.

We describe DSE methodology and calculate  the propagators in the next section~\myref{dsee},
the non-local condensate (NLC) and its response  to an external field is studied
in  section~\myref{nlc}. In section~\myref{eulerheisenberg} we do the Euler-Heisenberg
type effective action for quarks with non-perturbative  DSE propagators in external fields
and compare our results to the meson based evaluation.  We conclude in section~\myref{concl}.

\section{Dyson--Schwinger Equations\label{dsee}}
\subsection{Formulation of DSE with Quarks}
In this section we review the technique of obtaining quark and gluon
propagators in a self-consistent way. We use Fisher's  DSE technique
described in~\mycite{Fischer:2003zc}, and we show that the propagators
are reproduced by us in the case with quarks as well -- we have already
reproduced the gluodynamics sector in our previous study~\mycite{Zayakin:2009jz}.

\begin{figure}[h!] \begin{center}
\includegraphics[height = 4.cm, width=9cm]{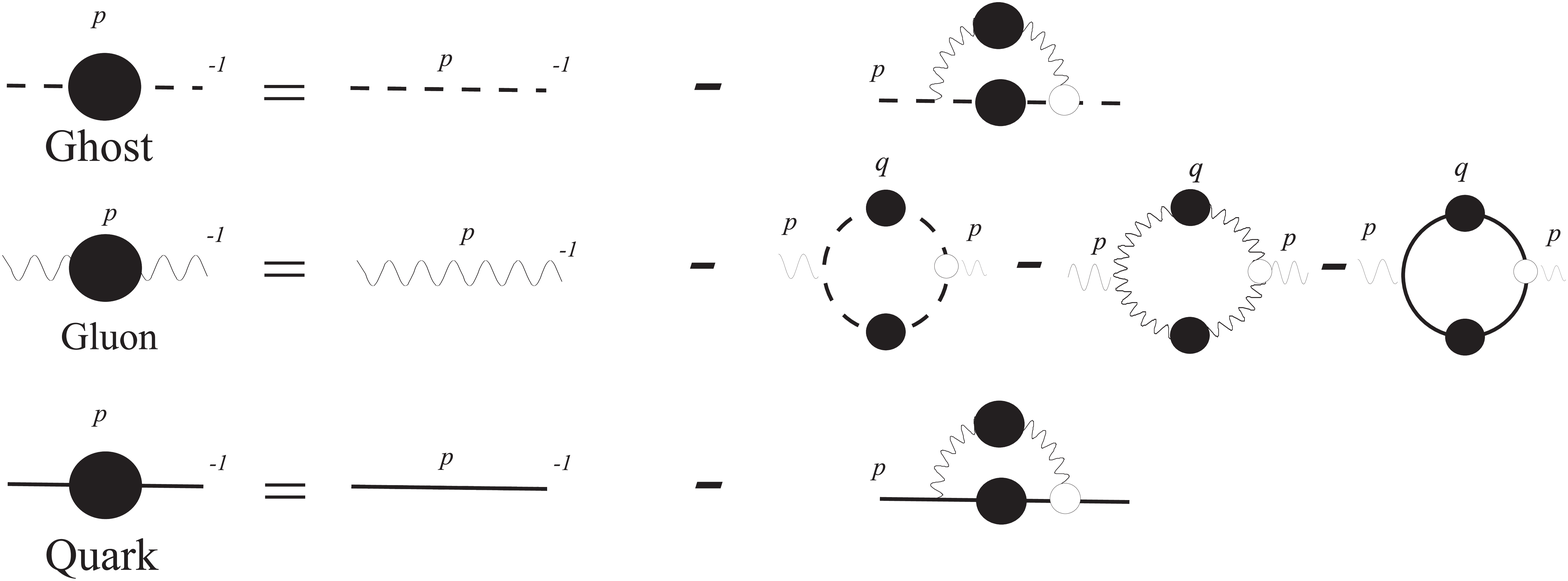}
\end{center}
\caption{Diagrammatic representation of DSE.
\label{dyson-diagr} }
\end{figure}

We apply in our work the Newton optimization method, based on the numerical procedure described
in~\mycite{Bloch:2003yu}.  We solve a system for ghost, gluon and quark propagators,
as shown in~\myfigref{dyson-diagr}. Propagator dressing is shown by bulbs,
and that of vertices -- by transparent bulbs.
We parameterize the gluon propagator in Landau gauge by the form-factor $F$, defined via relation
\beq
\displaystyle
D_{\mu\nu}^{F\,ab}(p) =
\delta^{ab}\left(g_{\mu\nu}-\frac{p_\mu p_\nu}{p^2}\right)\frac{ F(p^2)}{p^2+i\epsilon},
\eeq
and the ghost propagator is parameterized by the form-factor $G$
\beq
\displaystyle
D^{G\,ab}(p)=\frac{\delta^{ab}}
{p^2+i\epsilon}{ G(p^2)}.
\eeq
Quark propagator is defined as
\beq
S(p)=\frac{1}{A(p)}\frac{1}{\dirac{p}+M(p)}.
\eeq
Finding the scalar form-factors $F,G,A,M$ will yield non-perturbative information on the physical quarks and gluons.

DSE for this system can be written in the form:
\beq\displaystyle\label{dse}
\left\{\begin{array}{l}\displaystyle
\frac{1}{G(p^2)}-\frac{1}{G(\bar{\mu}_c^2)}=-\left(\Sigma(p^2)
-\Sigma(\bar{\mu}_c^2)\right),\\ \\ \displaystyle
\frac{1}{F(p^2)}-\frac{1}{F(\bar{\mu}_g^2)}
=-\left(\Pi(p^2)
-\Pi(\bar{\mu}_g^2)\right),\\ \\ \displaystyle
\frac{1}{A(x)}=1-\frac{\Pi_A(x)}{A(x)}
+\Pi_A(\bar{\mu}_g^2)\\ \\ \displaystyle
M(x)A(x)=M(\bar{\mu}_g^2)+\Pi_M(x)-
\Pi_M(\bar{\mu}_g^2)
\end{array}\right.
\eeq
Here $\bar{\mu}_{g,c}$ the are points of subtraction,
$\bar{\mu}_c=0$, $\bar{\mu}_g=\bar{\mu}$, $\bar{\mu}$
is the limit of the interval $p^2\in (0,\bar{\mu}^2)$
in the momentum space where we solve DSE, coupling $g^2$
is meant to be taken at point $\mu$: $g^2(\bar{\mu}^2)$. Gluon vacuum polarization is
\beq
\Pi(p^2)=\Pi^{2c}(p^2)+\Pi^{2g}(p^2)+\Pi^{2q}(p^2),
\eeq
contributions of ghosts $\Pi^{2c}(p^2)$  and gluons $\Pi^{2g}(p^2)$ being
\beq
\begin{array}{l}\displaystyle
\Pi^{2c}(p^2)=N_c g^2\int  \frac{d^d q}{(2\pi)^d} M_0(p^2,q^2,r^2)G(q^2)G(r^2),\\  \\ \displaystyle
\Pi^{2g}(p^2)=N_c g^2\int  \frac{d^d q}{(2\pi)^4} Q_0(p^2,q^2,r^2)F(q^2)F(r^2),
\end{array}
\eeq
ghost self-energy is
\beq
\Sigma(p^2)=N_c g^2\int K_0(p^2,q^2,r^2)G(q^2) F(r^2) \frac{d^d q}{(2\pi)^d}.
\eeq
Quark self-energy is conveniently split into functions $\Pi_A$ and $\Pi_M$, given below:
\beq\displaystyle\label{quarkM}
\begin{array}{l}\displaystyle
\Pi_M=\frac{1}{3\pi^3} \int  d^4y
\left\{
\frac{\alpha(z)}{z(y+M^2(y))}
\frac{G(z)^{-2d-d/\delta}}{F(z)^d}\frac{1}{A(y)}
\right.
\\ \\  \displaystyle \left.\left[\frac{3}{2}\left(A(x)+A(y)\right)M(y)
+\frac{1}{2}\left(\Delta A(x,y) M(y)-\right.\right.\right.\\
  \left.\left.\left.\displaystyle -\Delta B(x,y)\right)\left(-z+2(x+y)-(x-y)^2/z\right)
+\right.\right.\\ \\ \displaystyle \left.\left.+\frac{3}{2} (A(x)-A(y))M(y)\Omega(x,y)(x-y)
\right]
\right\}
\end{array}
\eeq
and
\beq\label{quarkA}
\begin{array}{l}
\displaystyle
\Pi_A=\frac{1}{3\pi^3} \int d^4y
\left\{
\frac{\alpha(z)}{xz(y+M^2(y))}
\frac{G(z)^{-2d-d/\delta}}{F(z)^d}\frac{1}{A(y)}\right.\\ \\ \displaystyle
\left.
\left[\left(-z+\frac{x+y}{2}+\frac{(x-y)^2}{2z}
\right)\frac{A(x)+A(y)}{2}-\right.\right.\\ \\
\displaystyle \left.\left.
-\left(\frac{\Delta A (x,y)}{2}(x+y)+\Delta B(x,y)M(y)\right)\times\right.\right.\\ \left.\left.\times
\left(-\frac{z}{2}+(x+y)
-\frac{(x-y)^2}{2z}\right)+
\right.\right.\\ \\ \displaystyle \left.\left.+\frac{3}{2}\left(A(x)-A(y)\right)\Omega(x,y)
\left(\frac{x^2-y^2}{2}-z\frac{x-y}{2}\right)
\right]
\right\},
\end{array}
\eeq
yielding the last two equations of\myref{dse}.
Here auxiliary functions $\Delta A,\Delta B,\Omega, \Delta\Omega$ have been introduced:
\beq
\begin{array}{ccl}
\Delta A(x,y)&=&\displaystyle{\frac{A(x)-A(y)}{x-y}},\\[0.4cm]
B(x)&=&M(x)A(x),\\[0.2cm]
\Delta B(x,y)&=&\displaystyle{\frac{B(x)-B(y)}{x-y}},\\[0.4cm]
\Omega(x,y)&=&\displaystyle{\frac{x+y}{(x-y)^2+(M^2(x)+M^2(y))^2}}.
\end{array}
\eeq
The constructions\myref{quarkM},\myref{quarkA}, are taken
from\mycite{Fischer:2003zc}, we have fixed here a typo
originally present in Eq.\myref{quarkM}.
Parameter $d$ is related to the Ansatz for the quark-gluon
vertex that is used. There is no unambiguous way of choosing
this parameter, since there is no fully consistent way of
truncating DSE without violating some of the worthy properties
of the original full tower of equations, and we refer the
reader to\mycite{Fischer:2003zc} for a comprehensive discussion on that point.
Variable $z$ is a logarithmic variable
\beq
z=\ln\frac{p^2}{\mu^2},
\eeq
and scale $\mu$ is yet to be defined upon solving DSE from comparing
the obtained coupling $\alpha_{DSE}(z)$ to the known values of
Particle Data Group coupling $\alpha_{PDG}(p^2)$~\cite{Amsler:2008zzb} at point $M$:
\beq\label{normscale}
\alpha_{DSE}(\ln(M^2/\mu^2))=\alpha_{PDG}(M^2).
\eeq
The coupling constant $g^2/4\pi\equiv \alpha$
is expressed in terms of $G,F$
solely\mycite{vonSmekal:1997is,vonSmekal:1997vx}, as vertex
is finite in Landau gauge (at one-loop level)
\beq\label{alphDSE}
\alpha_{\mathrm{DSE}}(\ln(p^2))=
\alpha_{\mathrm{DSE}}(\bar{\mu})F(p^2)G^2(p^2).
\eeq
In our case we shall use varying scale fixing  point $M$
so that we can prove that our results are independent
of scale fixing point choice within the error margin of our procedure.

The kernels $M_0,K_0,Q_0$ are well-known in literature,
however to make the presentation self-contained we provide them here:
\beq\displaystyle
\begin{array}{rcl}
K_0(x,y,\theta)&=&\displaystyle{\frac{y^2 \sin ^4(\theta )}
    {\left(-2 \cos (\theta ) \sqrt{x y}+x+y\right)^2}},\\[0.6cm]
M_0(x,y,\theta)&=&\displaystyle{-\frac{y^2 \sin ^4(\theta )}
     {3 x \left(-2 \cos (\theta ) \sqrt{x y}+x+y\right)}},
\end{array}
\eeq

\beq\displaystyle
\begin{array}{r}\displaystyle
Q_0(x,y,\theta)=-\frac{1}{12 x \left(-2 \cos (\theta )
 \sqrt{x y}+x+y\right)^2}\times\hspace*{1cm} \\[0.6cm]
\displaystyle
  \left\{y \sin ^2(\theta ) \left[2 \cos (2 \theta )
\left(6 x^2+31 x y+6 y^2\right)-\right.\right. \\[0.4cm] \displaystyle
-12 x \cos (3 \theta )
   \sqrt{x y}+x y \cos (4 \theta )
-48 \cos (\theta ) \sqrt{x y} (x+y)-\\ [0.4cm]\displaystyle
  \left.\left.
-12 y \cos (3 \theta ) \sqrt{x y}+3
   x^2+27 x y+3 y^2\right]\right\}.\phantom{\hspace*{1cm}}
\end{array}
\eeq
Scalar variables $x=p^2$, $y=q^2$ are introduced; variable
$\theta$ is defined via $(p-q)^2=x+y-2\sqrt{xy}\cos\theta$.
We neglect the effects of non-trivial
dressing of the vertices, since these do not essentially
back-react the upon the IR structure of the propagators themselves.

\begin{figure}[tbh]
\centerline{\includegraphics[height = 3.2cm, width=8cm]{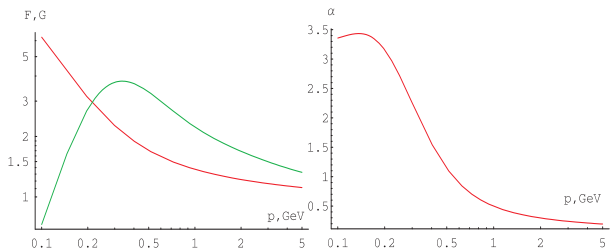}}
\caption{Ghost (dashed line) and gluon (solid line)
propagator form factors obtained in
DSE in Landau gauge; running coupling from DSE.
\label{prop}}
\end{figure}

To solve the Dyson--Schwinger equations we use the Ansatz~\mycite{Fischer:2002eq,Fischer:2002hna}:
\beq
\begin{array}{l}\displaystyle
F(z)=\left\{
\begin{array}{l}\displaystyle
\exp\left(\sum^{\bar{n}}_i a_i T_i(z)\right),\, z\in(\ln\epsilon,\ln\bar{\mu}^2),\\ \\ \displaystyle
F(\bar{\mu})\left(1+
\omega\log\frac{p^2}{\bar{\mu}^2}\right)^\gamma,
z>\ln\bar{\mu}^2,
  \\ \\ \displaystyle
A z^{2\kappa},z<\ln \epsilon,
\end{array}\right.\\  \\ \displaystyle
G(z)=\left\{
\begin{array}{l}\displaystyle
\exp\left(\sum^{\bar{n}}_i b_i T_i(z)\right),\,
   z\in(\ln\epsilon,\ln\bar{\mu}^2),\\  \\ \displaystyle
G(\sigma)\left(1+
\omega\ln\frac{p^2}{\bar{\mu}^2}
\right)^\delta,z>\ln\bar{\mu}^2,\\  \\ \displaystyle
B z^{-\kappa}, z<\epsilon.
\end{array}\right.,\\
\end{array}
\eeq
and similar Ans\"atze for $M(p),A(p)$.
Here $T_i$ are Tschebyschev polynomials, $a_i,b_i$
are unknown coefficients yet to be determined from
the numerical solution,  $\bar{n}$ is the number
of polynomials used (mostly $\bar{n}=30$ has been
used here, allowing precision of up to $10^{-10}$ for the
coefficients), $\delta=-9/44$,
$\gamma=-1-2\delta$, $\omega=11 N_c \alpha(\sigma)/(12\pi)$.
The IR scaling $\kappa$ is chosen to be the standard\mycite{Lerche:2002ep,Zwanziger:2001kw}
\beq
\kappa=0.59\label{kappa}
\eeq
for the case of Brown--Pennington
truncation with $\zeta=1$\mycite{Fischer:2002hna} 
(for discussion of meaning of $\zeta$ see \mycite{Brown:1988bm}), which is our case
($\zeta$ already set to its number value everywhere).
 Following\mycite{Alkofer:2008bs}, we employ renormalization
 constant $\mathcal{Z}_1$ redefinition,
so that no momentum dependence could possibly enter it, that is
\beq
\mathcal{Z}_1=\frac{G(y)^{(1-a/\delta -2a)}}{F(y)^{(1+a)}}
 \frac{G(y)^{(1-b/\delta -2b)}}{F(y)^{(1+b)}}.
\eeq
Again, following\mycite{Alkofer:2008bs} we choose
\beq
a=b=3\delta,
\eeq
which minimizes its momentum dependence. Renormalization
constant $\mathcal{Z}_1$ refers to the piece with a ghost
loop in vacuum polarization. The equations are solved by using
Newton's method,  described for this particular
application by Bloch~\mycite{Bloch:2003yu}.
 The results of the solution are propagator form factors
$F,G$, shown in\myfigref{prop} on the left,
the IR behavior of the propagators
corresponds to the standard ghost enhancement and gluon suppression.
The coupling $\alpha$ obtained from DSE \myref{alphDSE}  is shown
on the right in\myfigref{prop}. We note here that
the IR fixed point seen in the Figure is
\beq
\alpha(0)\approx 3
\eeq
for $N_c=3$, which is consistent with the up-to-date Dyson--Schwinger results reported
by other groups\mycite{Alkofer:2008bs,Huber:2007kc}.

\begin{figure}[tbh]
\begin{center}
\includegraphics[height = 4.5cm, width=8cm]{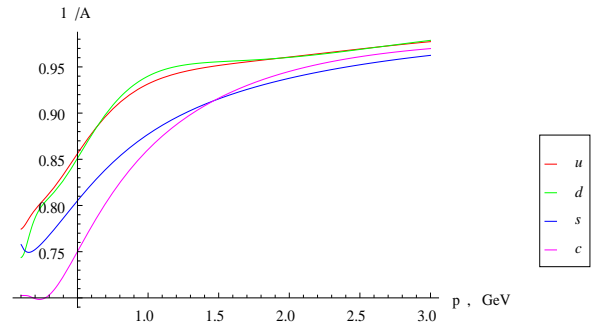}
\end{center}
\caption{Wave-function (propagator) form factors $A(p)$
for flavors $u\mbox{(red)},\, d\mbox{(green)},
s\mbox{(blue)},\,c\mbox{(magenta)}$ (lines from top to bottom at $p=0.5$ GeV.
\label{prop1}}
\end{figure}

Quark wave-functions were obtained for  one quark at a time
solving a in a selfconsistent way the DSE, i.e. these
are unquenched quarks. They are quite similar to quenched
approximation where quark DSE is solved for given glue DSE solution
(quenched approximation) The wave function form factors are shown
in\myfigref{prop1}. Wave-function form factors become perturbatively unity;
within an error margin they are no more distinguishable in the UV,
although they exhibit different and non-trivial behavior in the IR.

\begin{figure}[tbh]
\begin{center}
\includegraphics[height = 4.5cm, width=8cm]{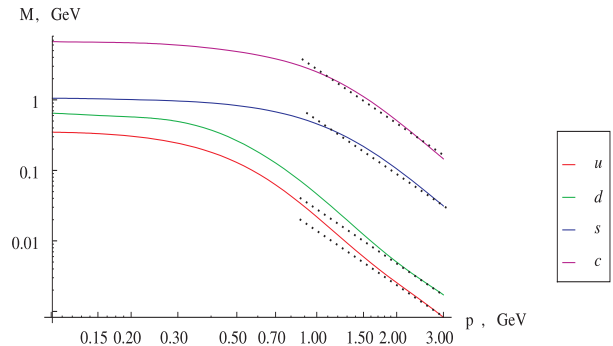}
\end{center}
\caption{Quark mass $M(p)$ for flavors $u,d,s,c$.
Punctured parallel tangent lines demonstrate that
anomalous dimension is mass-independent.
\label{prop2}}
\end{figure}

The quark masses  are shown in\myfigref{prop2}.
Physically it is important that UV anomalous dimensions of all
the quarks are rendered the same in\myfigref{prop2} , which confirms validity of
the procedure. This can be seen from the dashed parallel lines
in\myfigref{prop2}
In general, in this Section, we confirm all the current knowledge
on the DSE with quarks. We improve the numerical convergence by
smoothing numerical cut-off on integrals by superposing varying
limits, which procedure  removes Fourier transform `echos' from the
results.

\section{Non-local Condensate\label{nlc}}
\subsection{Dependence on mass of condensate shape}
In this Section we calculate the non-local condensate omitting the Wilson line,
study its behavior under external fields and compute the vacuum response
due to presence of non-local condensates to external fields.

The nonlocal condensate-related vacuum expectation value (C-VEV)
\beq
C_0(x)=\langle\bar{\psi}(x)\psi(0)\rangle,
\eeq
where local condensate $C(0)$ satisfies $ C(0)=C_0(0)$, can be related to propagator as
\beq
C_0(x)=\frac{1}{(2\pi)^4}N_c\sum_i^{N_f}\int d^4p 
 \frac{e^{ipx} }{A_i(p)}\frac{4M_i(p)}{p^2+M_i^2}- (\mbox{PT})
\eeq
However, separation the perturbative (PT)  part from the non-perturbative   propagator
is not well-defined. Moreover, some argue that the non-perturbative procedure
is producing only the non-perturbative quark propagator and there is no PT subtraction needed. 
We do not have a good argument to support this reasoning, or, alternatively,
a PT part subtraction, thus we follow the former approach. This  also does
not introduce additional procedure ambiguity.  Accordingly,
it should be remembered when evaluating  our results
that the full non-perturbative understanding of QCD
vacuum cannot be reached on grounds of   Dyson-Schwinger
equations alone, without applying additional resummation procedures,
e.g. ESSZ-resummation\mycite{Zayakin:2009jz}.
For this reason   our results should be treated as a first qualitative estimate,
and not yet as exact predictions.

\begin{figure}[h]
\begin{center}
\includegraphics[height = 4.5cm, width=8cm]{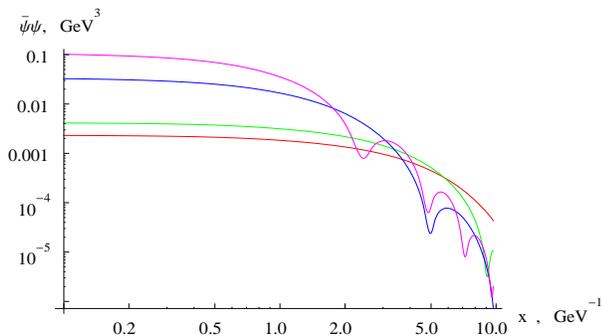}
 \end{center}
\caption{Non-local condensate for $u,d,c,s$ quarks.
\label{condensateUD}}
\end{figure}

We think that despite any of the above shortcomings the results  we obtain are
surprizing. We show  the   C-VEV in \myfigref{condensateUD},
where we see from top to bottom (at $x\to 0$)   beginning with the heavy quark
 $\langle \bar{c}(x) c(0)\rangle$,
 $\langle \bar{s}(x) s(0)\rangle$
 $\langle \bar{d}(x) d(0)\rangle$,
 $\langle \bar{u}(x) u(0)\rangle$,
and last the lightest $u$ quark. Numerical difficulties prevent us from
reaching higher mass than $500$ MeV for charm (at scale of 2 GeV).
Non-local condensate exhibits some oscillatory behaviour within
$2<x<10$ GeV$^{-1}$. We believe that these C-VEV oscillations
 are due to numeric uncertainty, but their
persistence and appearance when scale matching of quarks to glue and
ghost occurs suggest that further study of this phenomenon is needed,
and thus we show these results with C-VEV reaching to 2fm distance.
At this large distance the sequence of the C-VEV has reversed with
smallest quark mass leading to largest value of C-VEV.

\subsection{Local quark condensate and quark virtuality dependence on mass}
The standard wisdom\mycite{Shifman:1978bx} about condensate
 dependence on mass for heavy quarks is
\beq\label{vainshtein}
\langle\bar{q}q\rangle=-\frac{1}{12 m_q}\langle\alpha G^2\rangle.
\eeq
This relation is usually derived from requiring continuity between
heavy and light quarks' properties, imposed at the scale of about
 $0.2$ GeV. The behavior of our propagators  and wave functions
is continuous, yet the dependence on mass we observe is completely different.
Another well regarded relation is\mycite{Ioffe:2002ee}
\beq
\langle\bar{s}s\rangle\sim 0.8 \langle\bar{u}u\rangle.
\eeq

Note that in our evaluation the local condensate is independent
of Wilson line integral and thus
our results for $x\to 0$ while still PT subtraction dependent are more
secure. For $c$ and $s$ quarks  the values one sees in\myfigref{condensateUD}.
are considerably larger than expected. Moreover, we find that our
condensates increase with mass and does not decrease, as was expected
based on above estimates.

\begin{figure}[h]
\begin{center}
\includegraphics[height = 4.5cm, width=8cm]{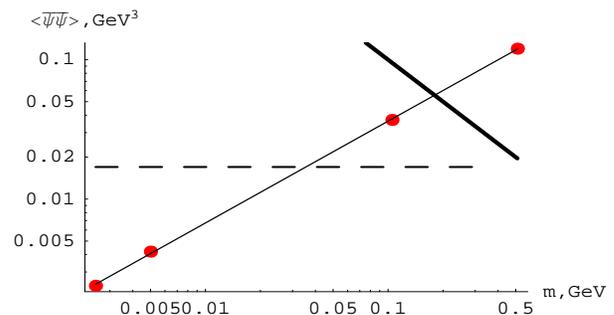}
 \end{center}
\caption{Local condensate mass dependence. red dots are DSE
results, thick line -- estimate \myref{vainshtein} for heavy quarks,
dashed line -- standard $\langle\bar{u}u\rangle$ value,
thin line -- power-law approximation\myref{power}.
\label{condensateonmass}}
\end{figure}

The values of condensate is fitted surprisingly well by a simple power law
\beq\label{power}
c(m)=0.2 \mathrm{GeV}^3\left(\frac{m}{\mbox{1GeV}}\right)^{0.73},
\eeq
not at all expected from any qualitative QCD model we know.
Mass dependence of condensate is illustrated in\myfigref{condensateonmass}.
The  dashed line is the expected light quark value, the thick line the
$c,s$ expectations of Eq.(\ref{vainshtein}). Note that these results are
obtained by considering one quark at a time and solving selfconsistently
 DSE (unquenched single quarks).

 It seems that with
increasing mass quarks can probe better the non-local glue vacuum fluctuations and
thus their response strength increases. The non-locality of glue vacuum structure
is usually not considered in the qualitative condensate models. However,
there is no argument we can present to align
the light  quark local condensate as function of $m$ with
heavy quark condensate.  We also note that
when dealing with realistic quarks, their
physical magnetic moments must be taken into account.
However this effect diminishes with quark mass and cannot
explain the heavy quark condensate behavior.

In another attempt to understand this strange behavior,
one could suggest that heavy quarks are worse represented
by Dyson--Schwinger equations since they tend to decouple
and thus one-loop approximation becomes almost free,
but at higher loops they might become again important,
 thus yielding DSE approach invalid.
However, this explanation is not valid, since comparison of
quenched approximation to the unquenched shows very little
difference between the two. Thus the issue of condensate
dependence on mass in DSE scheme presented here remains an open question.

Should this behavior be true, this strong dependence on mass of light quark condensate
would deeply impact  the chiral model analysis of quark masses,
where a cornerstone assumption
is that  light quark condensates have equal value.

\begin{figure}[tbh]
\begin{center}
\includegraphics[height = 3cm, width=5.5cm]{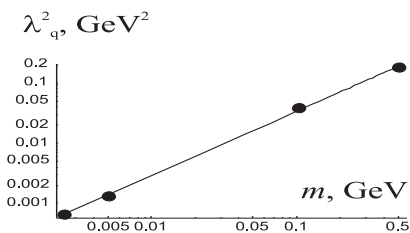}
\end{center}
\caption{Quark virtuality dependence on quark mass.
\label{virtu}}
\end{figure}

Quark virtuality dependence on mass is given in the table \ref{condensatechar}  below
and is shown in\myfigref{virtu}. We note the highly regular behavior, following the fit
\beq
\lambda^2_q=0.39 {\rm GeV}^2 \left(\frac{m_q}{1 {\rm GeV}}\right)^{1.07}
\eeq
shows in\myfigref{virtu} For comparison recall that   virtualities
$\lambda^2_{u,d}=0.7\mathrm{GeV}^2,
\lambda^2_{s}=1.6\mathrm{GeV}^2$ were reported\mycite{Zhou:2008zzl},
as discussed in Section 1. Recall also  $\lambda_q^2\approx 0.4\pm 0.1 \mathrm{GeV}^2$~\mycite{Chernyak:1983ej} and
$\lambda_q^2\sim 1.2\mathrm{GeV}^2$~\mycite{Shuryak:1988kd}.

\subsection{Condensate response to external field}
\begin{figure}[tbh]
\begin{center}
\includegraphics[height = 2.5cm, width=2.5cm]{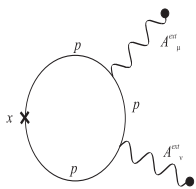}
\end{center}
\caption{Diagram describing condensate sensitivity to field.
\label{varcond}}
\end{figure}

\begin{figure}[tbh]
\begin{center}
\includegraphics[height = 4.5cm, width=8cm]{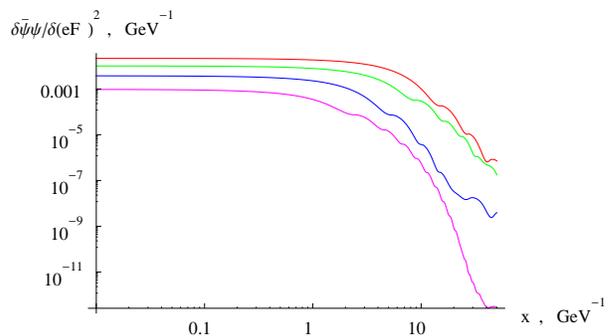}
\end{center}
\caption{Factor $f_1(x)$ as function of distance, describing the nonlocality of condensate sensitivity to field.
\label{dcondensate}}
\end{figure}

We can also establish the character of condensate dependence
on the external field. Considering a diagram\myfigref{varcond},
we derive the $F^2$-order term in the non-local condensate
\beq
\langle\bar{\psi}\psi\rangle_F=\langle\bar{\psi}\psi\rangle_0+
F^2 f_1(x)-F_{\nu\alpha} F^\alpha_\mu\frac{\partial^2 }{\partial_\nu\partial_\mu}f_2(x),
\eeq
where moments $f_1$, $f_2$ are
\beq
\begin{array}{l}\displaystyle
f_1(x)=\frac{1}{(2\pi)^4}\int \frac{e^{ipx} d^4 p}{A^3(p)}\frac{(-8)m(p)}{(p^2+m^2(p))^3},\\ \\
\displaystyle
f_2(x)=\frac{1}{(2\pi)^4}\int \frac{e^{ipx} d^4 p}{A^3(p)}\frac{(-16)m(p)}{(p^2+m^2(p))^4}.
\end{array}
\eeq
We notice here that not only the character of condensate dependence
on $x$ changes due to field switch-on, but it acquires anisotropy.
The function $f_1$ is shown in\myfigref{dcondensate}. It deserves
attention that smallest quark masses bring largest response to field,
which is quite reasonable. The resulting parameters are shown in the
table\myref{condensatechar}. It can be seen
from analysis of $f_1$ that already
fields of order of magnitude of
$10^{-1}$ GeV$^2$ may put the
local condensate to zero. This is comparable to the prediction of critical fields for condensate \beq
F_{\rm cr}=\frac{m_\pi^2}{\log 2}
\eeq
by Smilga and Shushpanov\mycite{Shushpanov:1997sf}.

\begin{table*}
\begin{flushleft}
\caption{\label{condensatechar}
Main characteristics of condensate: local amplitude $C(0)$, virtuality $\lambda^2$, condensate amplitude variation $\frac{\delta C(0)}{\delta F^2}$, virtuality variation $\frac{\delta \lambda_q^2}{\delta F^2}$, infrared exponent $a$ ($\langle\bar{\psi}(x)\psi(0)\rangle\sim e^{-ax}$), variation of the infrared exponent $\frac{\delta a}{\delta F^2}$. Mass value $m=0.51$ in the fourth line is not a misprint against the expected $m=1.27$ GeV, but was the largest mass at 2 GeV scale available to us.
}\end{flushleft}
$$
\begin{array}{|l|l|l|l|l|l|l|l|}\hline
m_q,\gev&q&C(0),\mbox{GeV}&\lambda^2_q,\gev^2
&\frac{\delta C(0)}{\delta F^2},\gev^{-1}&
\frac{\delta \lambda_q^2}{\delta F^2},\gev^{-2}&a,\gev&\frac{\delta a}{\delta F^2}, \gev^{-3}\\ \hline
 0.0025 & 2/3 & 0.00239 & 0.00066 & 0.037 & 0.00082 & 0.40 & 0.20 \\ \hline
 0.005 & 1/3 & 0.0042 & 0.0013 & 0.023 & 0.00094 & 0.65 & 0.20 \\ \hline
 0.105 & 1/3 & 0.037 & 0.039 & 0.015 & 0.0017 & 1.04 & 0.22 \\ \hline
 0.51 & 2/3 & 0.12 & 0.18 & 0.0064 & 0.0015 & 1.04 & 0.37 \\ \hline
\end{array}
$$
\end{table*}
Long-distance
correlations will be even more sensitive
to fields, since $f_1$ decreases
slower than condensate itself, thus making pion
wave-function a nice candidate for
analysis in an external field.

\section{Effective Action due to Condensates\label{eulerheisenberg}}
\begin{figure}[tbh]
\begin{center}
\includegraphics[height = 4.5cm, width=8cm]{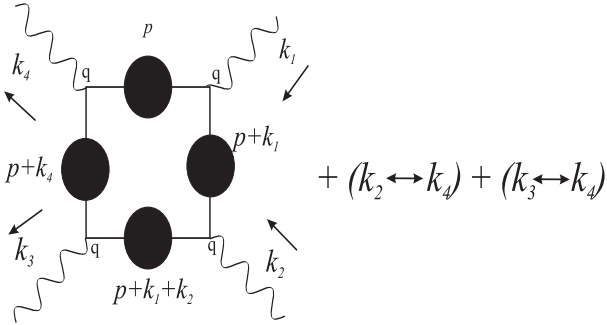}
 \end{center}
\caption{Leading nonlinear term in the Euler--Heisenberg effective action, $k_i$ are incoming momenta, $q$ quark charges.
\label{PhotonPhoton}}
\end{figure}

One of the simplest nonlinear processes of QED is photon-photon scattering,
shown in\myfigref{PhotonPhoton}. In the language of Euler--Heisenberg effective
action, the following term is responsible for this kind of processes
\beq\label{euler}
\begin{array}{l}
\mathcal{L}=a (F_{\mu\nu} F^{\mu\nu})^2+
  F_\mu^\nu F_\nu^\lambda F_\lambda^\rho F_\rho^\mu=\\ \\
  = A(F_{\mu\nu} F^{\mu\nu})^2 +B(F_{\mu\nu} \tilde{F}^{\mu\nu})^2.
\end{array}
\eeq
Coefficients $a,b$ are in case of QED
\beq
\begin{array}{l}\displaystyle
a=-\frac{\alpha^2}{36 m^4},\\  \\\displaystyle
b=\frac{7\alpha^2}{90 m^4},
\end{array}
\eeq
and $A,B$ are linearly related to them: $A=a+b/2, B=b$. We shall calculate
now these coefficients for condensate contribution of QCD vacuum
into QCD-related photon-photon scattering. We shall see at the end
that the contribution is larger than expected,  compared to standard (perturbative)
contribution due to hadrons. However, the magnitude of the effects
we find is very small compared to what is experimentally accessible today,
and in the foreseeable future, in the domain of intense laser physics.

Strictly speaking, when dealing with realistic quarks, their
physical magnetic moments must be taken into account. In the effective action
quark magnetic moments would invoke a contribution of the type
$\mu_q^4 F^{\mu_1\nu_1}F^{\mu_2\nu_2}F^{\mu_3\nu_3}F^{\mu_4\nu_4}
\tr[\sigma^{\mu_1\nu_1}\sigma^{\mu_2\nu_2}
\sigma^{\mu_3\nu_3}\sigma^{\mu_4\nu_4}]$. Noting that the outcome might
be  non-negligible, we leave this contribution
aside,  since it requires a  serious modification of the
DSE solution scheme and complicated issues of truncation validity.

To achieve the result we calculate the diagram\myfigref{PhotonPhoton}
with propagators obtained in the previous section, which are
responsible for condensates. We do not separate the condensate and the
free terms at the level of each propagator, but rather do the
full diagram with the full propagators, and then compare to the
perturbative terms ($a,b$ already given above, multiplied by
respective quark charges). As momentum dependence of the full
diagram on $s,t,u$ invariants would be known only numerically
as a result of a calculation, containing numerical data for
propagators, we use the following trick. We work out the
scattering amplitudes $M(e_1,e_2,e_3,e_4)$, given as
\beq
\begin{array}{l}
\tilde{M}_{\mu\nu\lambda\rho}(k_1,k_2,k_3,k_4)
=e^4\int \frac{d^4 p}{(2\pi)^4}\tr \left[S(p)\gamma^\mu S(p+k_1)\gamma^\nu \right.\\ \\
\displaystyle  \left.\hspace{0.5cm}S(p+k_1+k_2)\gamma^\lambda S(p+k_4)\gamma^\rho\right].
\end{array}
\eeq
For the scattering amplitudes at small values of photon frequencies
$\omega$ we extract the coefficient at the $\omega^4$ term:
\beq
\tilde{M}_{\mu\nu\lambda\rho}(k_1,k_2,k_3,k_4) e_1^\mu e_2^\nu e_3^\lambda e_4^\rho =
M_0+\omega^4 \alpha^2 M(e_1,e_2,e_3,e_4),
\eeq
for two specific sets of polarization vectors, namely, $(e_{1\bot},e_{2\bot},e_{3\bot},e_{4\bot})$
and $(e_{1\|},e_{2\|},e_{3\bot},e_{4\bot})$, ($e_{i\bot}$ denotes polarization orthogonal
to reaction plane, and $e_{1\|}$ polarization in reaction plane), at specific values
for $\theta$ (namely, forward scattering $\theta=\pi$).
These can be expressed as following scalar integrals
\beq
\begin{array}{l}\label{dsch}\displaystyle
M(e_{1\bot},e_{2\bot},e_{3\bot},e_{4\bot})=
\\ \\ \displaystyle=\int_0^\infty \frac{32 p^3 dp }{15 \left[p^2+M(p)^2\right]^8 A(p)^3}
  \left[19 p^8+75 M(p)^2 p^6 -\right.\\ \\ \left. -10 M(p)^4 p^4-330 M(p)^6 p^2+30
   M(p)^8\right]\\ \\
\end{array}
\eeq
\beq
\begin{array}{l}
\displaystyle
M(e_{1\|},e_{2\|},e_{3\bot},e_{4\bot})=\\  \\ \displaystyle
=-\int_0^\infty \frac{32 p^3 dp }{15 \left[p^2+M(p)^2\right]^8 A(p)^3}
\left[7 p^8-25 M(p)^2 p^6-\right.\\  \\ \left.
-40 M(p)^4 p^4+60 M(p)^6 p^2-30
   M(p)^8\right].
\end{array}
\eeq
Polarization vectors have been
\beq
\begin{array}{l}
e_\|=\{0,0,1,0\},\\
e_\bot=\{0,0,0,1\},
\end{array}
\eeq
with center-of-mass kinematics
\beq
\begin{array}{l}
k_1=\omega\{1,1,0,0\},\\
k_2=\omega\{1,-1,0,0\},\\
k_3=\omega\{1,\cos\theta,\sin\theta,0\},\\
k_4=\omega\{1,-\cos\theta,-\sin\theta,0\}.
\end{array}
\eeq
In the expansion we used the fact that $\omega$ is believed to be small, therefore, all non-perturbative momentum-dependent factors ($M(p), A(p)$) are taken at the point $p$.

On the other hand, the coefficients $M(\dots)$ are known from\myref{euler} by direct analysis
\beq
\begin{array}{l}\label{ahiezer}
M(e_{1\bot},e_{2\bot},e_{3\bot},e_{4\bot})=64(2a+b),\\ \\
M(e_{1\|},e_{2\|},e_{3\bot},e_{4\bot})=16(4a+b).
\end{array}
\eeq
Thus a simple comparison of \myref{dsch} and \myref{ahiezer} yields values for $a,b$ and $A,B$. They are shown in Table\myref{abAB}.
\begin{table}\begin{flushleft}
\caption{Coefficients $a,b,A,B$ of non-linear terms in the effective action. The ``perturbative'' line shows for comparison the coefficients $a_0,b_0,A_0,B_0$ for mass $m$ in the loops which can be thought of approximately as $\Lambda\sim 300 \mathrm{GeV}$; our results are shown as dimensionless ratios against $a_0,b_0,A_0,B_0$. Quarks charges $q_i=2/3,1/3$ are included into coefficients.
\label{abAB}}\end{flushleft}
\begin{equation*}
\begin{array}{|l|l|l|l|l|}\hline
&a_0&b_0&A_0&B_0\\ \hline
PT&-\frac{1}{36m^4}&\frac{7}{90 m^4}&\frac{1}{90m^4}&\frac{7}{90m^4}\\ \hline\hline
{\rm flavor}&a/a_0&b/b_0&A/A_0&B/B_0\\ \hline
u& 0.07732 & 0.09317 & 0.1328 & 0.09317 \\ \hline
d& 0.00302 & 0.00337 & 0.00425 & 0.00337 \\  \hline
s& 0.00019 & 0.00022 & 0.0003 & 0.00022 \\  \hline
c& 0.00064 & 0.0007 & 0.00085 & 0.0007 \\
  \hline
\end{array}
\end{equation*}
\end{table}
This Table is quite instructive. First of all, the contributions
 are comparable with the expected hadronic ones. 
The range of the latter can be estimated roughly within
$1/90 m_\pi^4\dots 1/90 m_\rho^4\sim 40\dots 0.05$~GeV$^{-4}$.
Quarks with large bare masses yield less, as expected on general grounds.

Basing on the analysis of light quark properties we confirm the
claims that QCD vacuum cannot be probed
via non-linear QED effects until fields of hadron scales
$m_h^2\sim (0.1 \gev)^2$ are reached, actually on scales
several times below typical hadronic scale the effect can
be felt as contribution to the naive QED estimate of
photon-photon scattering.

We believe that the issue of
heavy quarks, which have yielded several
inconsistencies in the suggested scheme,
could be addressed in our scheme once more
effort is devoted to understand the solutions
of DSE. Moreover, to arrive at realistic results we
need to consider schemes with all quarks participating in the
solution of DSE equation, and allowing for vertex correction
which implements magnetic moment for light quarks.

\section{Conclusion\label{concl}}
Solving one quark-gluon-ghost Dyson-Schwinger equations, we have obtained
the quark non-local condensate and quark virtuality  as functions of quark mass.
The mass dependence of the condensate
  disagrees with current qualitative wisdom. We could not
find an explanation for why this is the case. The growth of the
quark condensate with $m^{0.73}, m<500\,\mathrm{MeV}$
implies a significant difference between
all mass condensates above and beyond any expectations. If confirmed,
this result would have considerable impact on hadron phenomenology.
For example, the difference in $u$-quark and $d$-quark mass and thus
implied difference in quark condensate leads to
considerable change in chiral analysis of quark masses.

Regarding the influence of an external field on the  condensate we
predict that fields of order of magnitude of order of magnitude of
$10^{-1}$  GeV$^2$ can actually destroy local condensate,
and even smaller fields can destroy non-local $x$-dependent
condensate at $x\neq 0$. This result may have direct impact  the
pion wave function in external fields. As recently shown
by Pimikov, Bakulev and Stefanis\mycite{Pimikov:2009mq}, it is the non-locality of the
condensates that is the key point for inclusion of the non-perturbative
contributions to the pion form factor. Thus our results on condensates
immediately drive the dynamics of pion wave function in external fields.

In addition we predict that light quarks yield important non-perturbative
contributions into the photon-photon scattering amplitude, which are comparable
with the corresponding perturbative contributions based on loops with light mesons.
We stress that this dynamics is essentially condensate-driven,
and particularly by its non-locality. Our present non-perturbative evaluation
suggests that   the critical field, above which the non-linear QCD-QED effects can be seen, is
several times lower than the typical hadronic scale. Even so, the experiments to
probe the QCD vacuum with intense laser fields are beyond the foreseeable future.

We have outlined in the text an opportunity for further advance which
must first focus on the resolution of the mass dependence of quark condensate and
better understanding of the related virtuality.  The relatively large effects
which external fields can impart on the QCD vacuum must be confirmed in the
context of such an improved theoretical framework.

\section*{Acknowledgments}
We thank  R.Alkofer, A.Bakulev, S.J.Brodsky, C.Fischer
J.Pawlowski, A.Pimikov, M.I.Polikarpov,   L. von Smekal
for discussions. We are specially grateful to C. Fischer who supplied us with his data on
propagators, thus greatly simplifying the tedious process of solving DSE.

AZ  thanks   F.Gubarev, D.Habs, A.Maas, Yu.M.Makeenko, S.Mikhailov,
M.Pak, V.Shevchenko, D.V.Shirkov,  V.I.Zakharov, K.Zarembo for
in depth discussions on related questions
 of non-perturbative QCD. AZ also thanks the  Organizers of
the Tenth Workshop on Non-Perturbative Quantum Chromodynamics
at l'Institut Astrophysique de Paris,
June 8-12, 2009, for creative scientific atmosphere.

JR thanks Jan Pawlowski for sharing Fish and Chips on the way
to  the Cairns QCD Topical Workshop which started work on this project.
He further thanks Herrn Dietrich Habs for his interest and  enthusiastic
Bavarian hospitality at the Munich Centre of Advanced Photonics -- DFG excellence
cluster in Munich.

This work was supported  by the DFG Cluster of Excellence MAP (Munich Centre of
Advanced Photonics). AZ was in part supported by the RFBR grant 07-01-00526.
JR work was supported  by a grant from the U.S. Department of Energy  DE-FG02-04ER41318.



\providecommand{\noopsort}[1]{}\providecommand{\singleletter}[1]{#1}%

\end{document}